\documentclass[a4paper,fleqn,usenatbib]{mnras}

\usepackage[T1]{fontenc}
\usepackage{ae,aecompl}
\usepackage{graphicx}	
\usepackage{amsmath}	
\usepackage{amssymb}	

\title[Theory of transmission spectra revisited]{The theory of transmission spectra revisited: a semi-analytical method for interpreting WFC3 data and an unresolved challenge}

\author[Heng \& Kitzmann]{
Kevin Heng$^{1}$\thanks{E-mail: kevin.heng@csh.unibe.ch (KH)}
and Daniel Kitzmann$^{1}$\thanks{Email: daniel.kitzmann@csh.unibe.ch (DK)}
\\
$^{1}$University of Bern, Center for Space and Habitability, Sidlerstrasse 5, CH-3012, Bern, Switzerland\\}

\date{Accepted 2017 June 8. Received 2017 June 5; in original form 2017 February 7.}

\pubyear{2016}

\begin{document}
\label{firstpage}
\pagerange{\pageref{firstpage}--\pageref{lastpage}}
\maketitle

\begin{abstract}
The computation of transmission spectra is a central ingredient in the study of exoplanetary atmospheres.  First, we revisit the theory of transmission spectra, unifying ideas from several workers in the literature.  Transmission spectra lack an absolute normalization due to the a priori unknown value of a reference transit radius, which is tied to an unknown reference pressure.  We show that there is a degeneracy between the uncertainty in the transit radius, the assumed value of the reference pressure (typically set to 10 bar) and the inferred value of the water abundance when interpreting a WFC3 transmission spectrum.  Second, we show that the transmission spectra of isothermal atmospheres are nearly isobaric.  We validate the isothermal, isobaric analytical formula for the transmission spectrum against full numerical calculations and show that the typical errors are $\sim 0.1\%$ ($\sim 10$ ppm) within the WFC3 range of wavelengths for temperatures of 1500 K (or higher).  Third, we generalize the previous expression for the transit radius to include a small temperature gradient.  Finally, we analyze the measured WFC3 transmission spectrum of WASP-12b and demonstrate that we obtain consistent results with the retrieval approach of \cite{k15} if the reference transit radius and reference pressure are fixed to assumed values.  The unknown functional relationship between the reference transit radius and reference pressure implies that it is the product of the water abundance and reference pressure that is being retrieved from the data, and not just the water abundance alone.  This degeneracy leads to a limitation on how accurately we may extract molecular abundances from transmission spectra using WFC3 data alone.  We suggest an approximate expression for this relationship.  Finally, we compare our study to that of \cite{g14} and discuss why the degeneracy was missed in previous retrieval studies.
\end{abstract}


\section{Introduction}
\label{sect:intro}

A substantial fraction of the measurements made of exoplanetary atmospheres takes the form of transmission spectra---scrutinizing the small change in the projected size of the exoplanet, across wavelength, as it transits its host star \citep{ss00}.  To interpret a transmission spectrum using atmospheric retrieval requires that we are able to solve the inverse problem robustly and efficiently: given the wavelength-dependent transit radius, $R(\lambda)$, we wish to infer the types and abundances of atoms and molecules that contribute to the opacity of the atmosphere.  Across a transmission spectrum, the relative size of spectral features is proportional to the pressure scale height, which depends on temperature, surface gravity and mean molecular mass.

Traditionally, one calculates the transmission spectrum by tracing a set of rays through the limb of the atmosphere \citep{brown01,h01,b03}.   Each ray passes through a transit chord with varying degrees of transparency or opaqueness.  By summing up the contributions from all of these chords, one may calculate the effective occulting area of the exoplanet at a given wavelength, $\pi R^2$.  Several studies describe the theory of transmission spectra \citep{ss00,brown01,h01,b03,fortney05,bs12,ds13,g14,v14,heng15,bs17}, but a unified analytical approach is missing from the literature.  Part of the motivation of the present study is to present such a unified approach that thoroughly explores the assumptions, caveats and degeneracies associated with the calculation of transmission spectra.  For the present study, we ignore the effects of refraction \citep{ss00,brown01,h01,b16} and multiple scattering \citep{brown01,robin17}.

\begin{figure}
\vspace{-0.1in}
\includegraphics[width=\columnwidth]{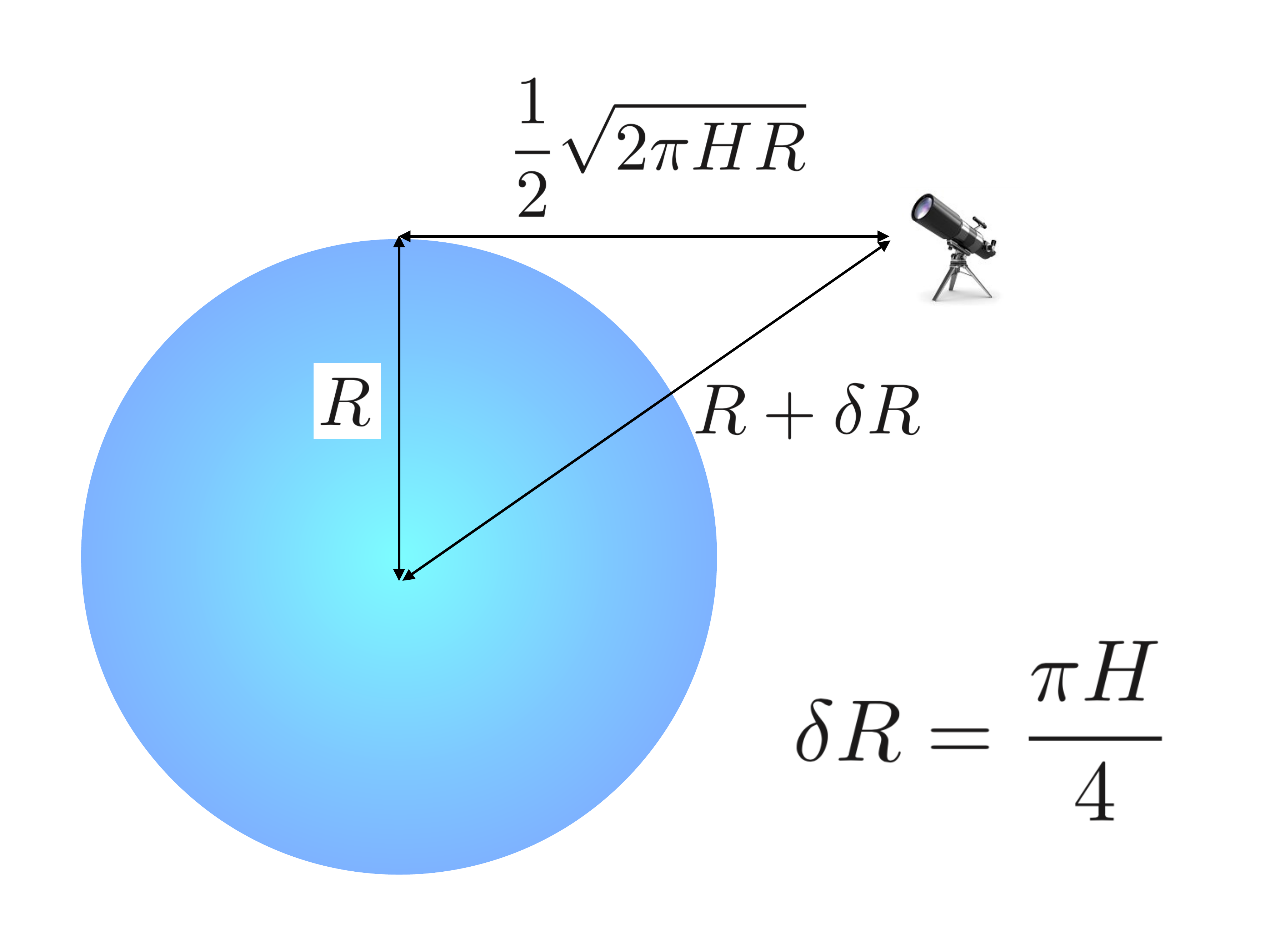}
\vspace{-0.1in}
\caption{Schematic illustrating the transit geometry for an isothermal atmosphere, which has an effective chord length of $\sqrt{2 \pi HR}$ corresponding to the transit radius \citep{fortney05}.  This works out to $\delta R = \pi H/4$, which is a change in pressure of a factor of $\exp{(\pi/4)} \approx 2.19$ across the transit chord.}
\label{fig:schematic}
\end{figure}

Transmission spectra have previously been shown to be insensitive to temperature variations \citep{brown01,h01}.  (In the present study, we will show that $R \propto \kappa^{\pm 1/b} \ln{\kappa}$ in the presence of a temperature gradient, where one may reasonably expect $b \gg 1$.  The opacity, $\kappa$, depends on the temperature.) Retrieval approaches for transmission spectra have assumed isothermal temperature-pressure profiles---either explicitly \citep{w15} or implicitly \citep{bs12}\footnote{\cite{bs12} used the analytical temperature-pressure profile of \cite{g10}, which is perfectly isothermal at high altitudes due to the underlying, simplifying assumptions made, i.e., that the absorption-, flux- and Planck-mean opacities are equal.}.  For an isothermal atmosphere, it has previously been shown that the transit chord has a length of $\sqrt{2 \pi H R}$, with $H$ being the pressure scale height \citep{fortney05}.  The transit geometry is such that $\sqrt{\pi H R/2}$, $R$ and $R + \delta R$ form the three sides of a triangle, from which we may calculate that $\delta R = \pi H/4$ (Figure \ref{fig:schematic}).  This in turn corresponds to a factor of $\exp{(\pi/4)} \approx 2.19$ for the change in pressure across the transit chord.  Thus, isothermal atmospheres are \textit{nearly} isobaric, which leads us to the reasoning that an analytical formula for an isothermal, isobaric atmosphere should adequately serve as an accurate fitting function to data.  Furthermore, transmission spectra should be insensitive to the pressure being probed as long as the temperature is high enough and pressure broadening is relatively unimportant at the altitudes being sensed.

The noise floor for the \textit{Wide-Field Camera 3} (WFC3) on the \textit{Hubble Space Telescope} is typically 50 ppm and does not reach 20 ppm, and the instruments on the \textit{James Webb Space Telescope} (JWST) are expected to have similar noise floors \citep{greene16} due to the shared heritage between the infrared detectors \citep{beich14}.  Thus, an isothermal, isobaric formula is deemed to be accurate enough if the average error associated with it is better than 20 ppm.  Certainly, the true noise floor for transmission spectra obtained using JWST awaits the procurement and analysis of real data.

For completeness, we also generalize the isothermal formula for the transit radius to include a (small) temperature gradient in the region of the atmosphere probed by transmission spectroscopy.  Validated analytical formulae are invaluable for studying degeneracies in the model without expending much computational effort.

An under-emphasized fact is that the normalization of a transmission spectrum depends on specifying a reference transit radius that is associated with a reference pressure.  Higher temperatures or a finite temperature gradient may be compensated by lower values of the normalization.  The value of the reference transit radius may be matched\footnote{We emphasize that this is the \textit{choice} of the modeller.  Theoretically, any transit radius may be chosen as a reference.} to the measured white-light radius, but this measurement is associated with an uncertainty.  Furthermore, the functional relationship between the reference transit radius and reference pressure is unknown.  The reference pressure cannot be extracted from the data and must be assumed.  It is typically set to 10 bar (e.g., \citealt{line13,k15}).  A goal of the present study is to show that there is a degeneracy between the reference transit radius, reference pressure and water abundance that cannot easily be overcome. 

\begin{figure}
\vspace{-0.2in}
\includegraphics[width=\columnwidth]{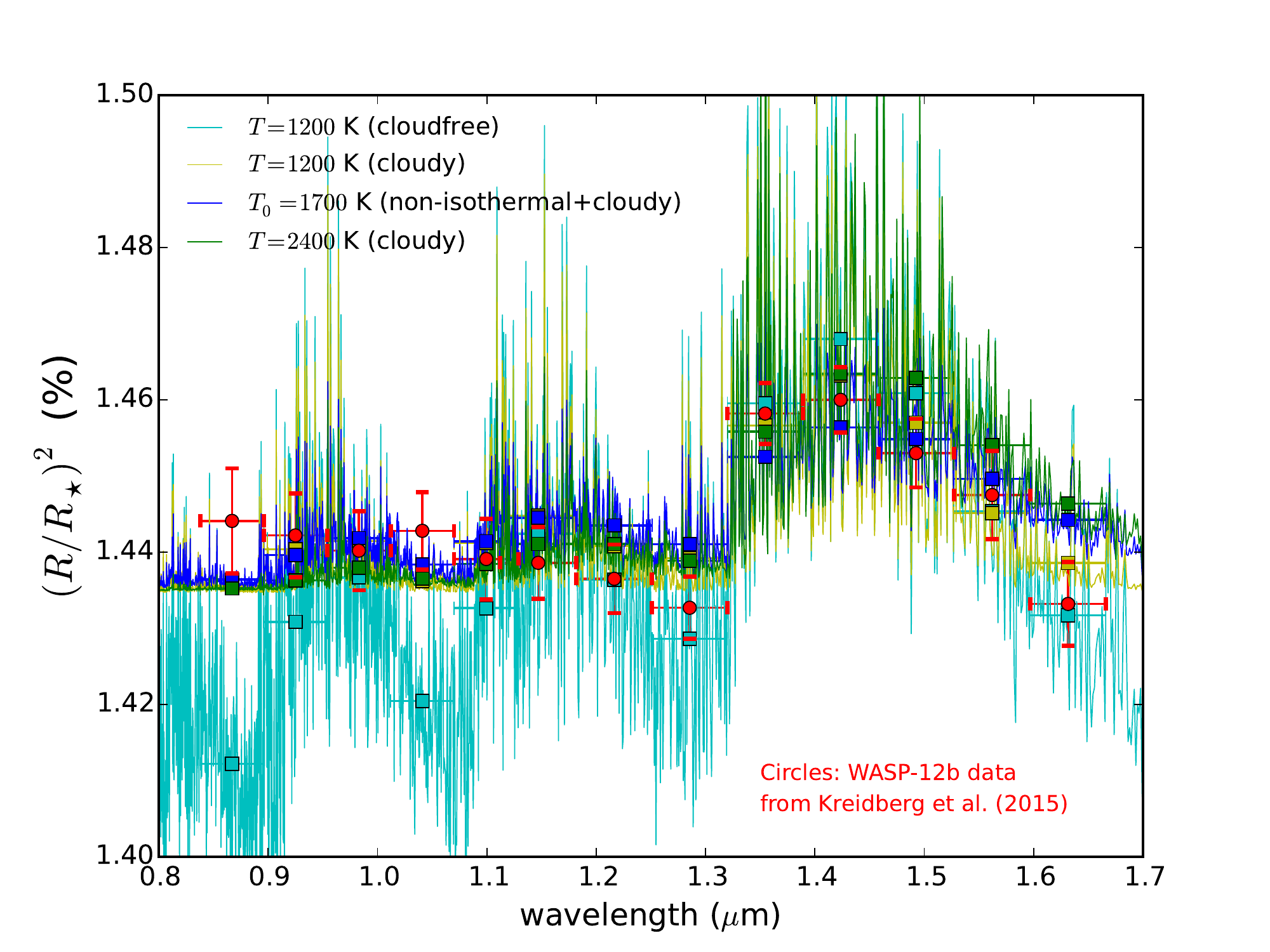}
\vspace{-0.1in}
\caption{Examples of model transmission spectra with different temperatures, cloud opacities and a finite temperature gradient.  The water volume mixing ratio is set to $10^{-3}$ for illustration.  See Table \ref{tab:wasp12b} and text for the parameter values assumed for these models.}
\label{fig:wasp12b}
\end{figure}

In Figure \ref{fig:wasp12b}, we show examples of model transmission spectra specialized to the wavelength range of WFC3 onboard the \textit{Hubble Space Telescope}, and over-plot the data of WASP-12b measured by \cite{k15}.  (The parameter values used are listed in Table \ref{tab:wasp12b}.)  It is apparent that the temperature, temperature gradient and degree of cloudiness are somewhat degenerate quantities that produce similar model transmission spectra.  The spectral features just blueward and redward of the 1.4 $\mu$m water feature break these degeneracies partially and allow for a unique fit to be obtained \textit{if the reference transit radius and reference pressure are fixed.}  Another goal of the study is to introduce a fast method for performing this fit to data using a validated analytical formula, while pointing out that \textit{our ignorance of the relationship between the reference transit radius and reference pressure plagues our ability to extract accurate values of the water abundance.}  We then suggest an approximate expression to relate the reference transit radius and reference pressure.

In \S\ref{sect:formalism}, we lay down the formalism for computing transmission spectra for both isothermal and non-isothermal atmospheres.  In \S\ref{sect:results}, we benchmark our isothermal, isobaric analytical formula and report on its accuracy.  We apply our formula to the case study of WASP-12b and discuss the degeneracies involved in data interpretation.  In \S\ref{sect:discussion}, we discuss the implications of our results.

\section{Formalism}
\label{sect:formalism}

\begin{table}
\centering
\caption{Parameter values used for WASP-12b models of transmission spectra in Figure \ref{fig:wasp12b}}
\label{tab:wasp12b}
\begin{tabular}{lcccc}
\hline
$T$ & $T_0$ & $T^\prime$ & $R_0$ & $\kappa_{\rm cloud}$ \\
\hline
(K) & (K) & (K km$^{-1}$) & ($R_{\rm J}$) & (cm$^2$ g$^{-1}$) \\
\hline
\hline
1200 & -- & -- & 1.790 & 0 \\
1200 & -- & -- & 1.786 & $2 \times 10^{-3}$ \\
-- & 1700 & 0.1 & 1.775 & $2 \times 10^{-3}$ \\
2400 & -- & -- & 1.711 & $4 \times 10^{-2}$ \\
\hline
\end{tabular}\\
Note: Following \cite{k15}, we assume $R_\star = 1.57 ~R_\odot$ and $P_0=10$ bar.  For illustration, we assume a negative temperature gradient in our $T_0=1700$ K example.  Our water opacities are computed at a pressure of 1 mbar.
\end{table}

\subsection{Order-of-magnitude expressions}

Consider a toy atmosphere with a constant number density $n$, containing a single molecule with an extinction cross section $\sigma$.  A single temperature $T$ describes this atmosphere.  The pressure scale height is $H = kT/mg$ with $k$ being the Boltzmann constant, $m$ being the mass of the molecule and $g$ being the surface gravity.  Since we expect $H \ll R$, the characteristic length scale in the system is the geometric mean of $H$ and $R$, which is $\sqrt{HR}$.  The chord optical depth associated with the transit radius is
\begin{equation}
\tau \sim n \sigma \sqrt{HR} \sim 1.
\end{equation}
If we re-arrange this expression and invoke the ideal gas law, we obtain an expression for the transit radius in terms of the opacity ($\kappa$; cross section per unit mass),
\begin{equation}
R \sim H \left( \frac{g}{P \kappa} \right)^2.
\end{equation}
Consider water to be the only opacity source in the atmosphere (since the WFC3 bandpass probes mostly water).  This expression already reveals the presence of a degeneracy between the pressure, temperature and abundance of water.  The central limitation is that there is one equation and three unknowns.

Setting $\tau \sim 1$ is reasonable but imprecise, because one needs to formally integrate over a range of chords with various values of $\tau$.

\subsection{Isothermal atmospheres}

We revisit the formalism for isothermal chord optical depths and transit radii, first explored by \cite{fortney05} and later expanded upon by \cite{ds13}, \cite{heng15} and \cite{bs17}.  Our intention is to clarify several aspects of the derivation and elucidate the assumptions made.

Consider that an observer records a transit radius of $R^\prime$.  A radial coordinate $r^\prime$ is defined such that $r^\prime=0$ is located exactly at $R^\prime$.  It follows that $r^\prime \approx x^2/2R^\prime$, where $x$ is the spatial coordinate along the sightline of the observer \citep{fortney05}.  For an isothermal atmosphere, we have $n = n^\prime \exp{(-r^\prime/H)}$, such that $n^\prime$ is the number density probed at $R^\prime$.  By evaluating $\int^{+\infty}_{-\infty} n \sigma ~dx$ and assuming that $\sigma$ may be taken out of the integral, we obtain $\tau = n^\prime \sigma \sqrt{2 \pi H R^\prime}$ \citep{fortney05}.  An order-of-magnitude estimate of $\tau$ misses the $\sqrt{2\pi}$ factor.  It is important to note that the characteristic length scale, $\sqrt{2 \pi HR^\prime}$, only appears when the integration is carried out formally from $-\infty$ to $+\infty$.

Now, we define a radial coordinate $r$ such that $r=0$ is located at the center of the exoplanet.  We rescale $r^\prime$ such that $r^\prime=0$ sits at $r=R_0$.  Let the number density associated with the reference transit radius ($R_0$) or pressure ($P_0$) be $n_0$.  It follows that $n^\prime = n_0 \exp{[-(r-R_0)/H]}$.  Since $R^\prime = R_0 + r^\prime \approx R_0$, the chord optical depth is, to a good approximation,
\begin{equation}
\tau = \tau_0 \exp{\left( - \frac{r-R_0}{H} \right)},
\label{eq:tau}
\end{equation}
which agrees with equation (S.1) of \cite{ds13} and equation (15) of \cite{bs17}.  The reference optical depth is
\begin{equation}
\tau_0 =  \frac{P_0 \sigma}{k T} \sqrt{2 \pi H R_0}.
\end{equation}
These considerations yield a relation between $\tau$ and $r$,
\begin{equation}
dr = - \frac{H d\tau}{\tau}.
\end{equation}

Let the effective thickness of the atmosphere, at a given wavelength, be $h$.  The transit radius is then $R = R_0 + h$.  The projected size of the exoplanet is $\pi R^2$, and it may also be expressed as $\pi R_0^2 + A$ with $A$ being the area of the annulus sitting above the reference radius \citep{brown01,ds13},
\begin{equation}
A = 2 \pi \int^{+\infty}_{R_0} \left[ 1 - \exp{\left(-\tau\right)} \right] r ~dr.
\end{equation}
Note that the integration is formally carried out to $+\infty$, even though the transit radii is effectively located at a finite distance from the center of the exoplanet.  Since we generally expect $A/\pi R_0^2 \ll 1$, the effective thickness of the atmosphere is
\begin{equation}
h = \frac{A}{2 \pi R_0}.
\end{equation}

By performing a change of coordinate from $r$ to $\tau$, we obtain
\begin{equation}
h = H \int^{\tau_0}_0 \left( \frac{1 - e^{-\tau}}{\tau} \right) \left[ 1 + \frac{H}{R_0} \ln{\left( \frac{\tau_0}{\tau} \right)} \right] ~d\tau.
\label{eq:h}
\end{equation}
As previously reasoned by \cite{bs17}, the term associated with $H/R_0$ is much smaller than the other term and may therefore be neglected, whereas \cite{ds13} kept both terms and showed that hypergeometric functions obtain from evaluating the smaller term.  Following \cite{bs17}, we use the identity in equation (10) of Appendix I of \cite{chandra},
\begin{equation}
E_1 = -\gamma - \ln{\tau_0} + \int^{\tau_0}_0 \frac{1 - e^{-\tau}}{\tau} ~d\tau,
\end{equation}
where $\gamma \approx 0.57721$ is the Euler-Mascheroni constant.  The quantity $E_1$ is the exponential integral of the first order with the argument $\tau_0$.  As $\tau_0 \rightarrow \infty$, we have $E_1 \rightarrow 0$.  Applying the preceding identity to equation (\ref{eq:h}) yields
\begin{equation}
h = H \left( \gamma + \ln{\tau_0} + E_1 \right).
\end{equation}
The extra $E_1$ term was not explicitly stated in equation (7) of \cite{ds13}, but is implicitly present in their equation (S.4), contrary to the claim of \cite{bs17}.

\cite{bs17} have previously interpreted $\tau_0$ to be associated with an optically thick surface such as a cloud deck.  In the current derivation, $\tau_0$ is simply associated with a reference pressure corresponding to an atmospheric layer that may---or may not---be chosen to be optically thick.  It is a natural outcome of the a priori unknown value of the pressure associated with the reference transit radius $R_0$.  We generally expect $\tau_0 \gg 1$, which means that the $E_1$ term vanishes and one ends up with an expression for the effective chord optical depth associated with the transit radius $R$,
\begin{equation}
\tau_{\rm eff} = \tau_0 \exp{\left(- \frac{h}{H} \right)} = \exp{\left(-\gamma \right)} \approx 0.56.
\end{equation}
As already remarked by \cite{ds13} and \cite{bs17}, this value of 0.56 is derived from first principles, whereas \cite{lec08} inferred it by using a graphical solution.

Finally, we state the expression for the transit radius assuming an isothermal atmosphere,
\begin{equation}
R = R_0 + H \left[ \gamma + \ln{\left( \frac{P_0 \kappa}{g} \sqrt{\frac{2 \pi R_0}{H}} \right)} \right].
\label{eq:isothermal}
\end{equation}
As expected, the transit radius depends linearly on the pressure scale height, but is a slowly varying function of the opacity \citep{brown01}.  The opacity is evaluated at the temperature $T$ and a pressure $P$ that is arbitrarily chosen such that pressure broadening is negligible.  However, since the opacity typically varies over many orders of magnitude, its overall effect on the transmission spectrum is comparable to that of $H$.

The formula in equation (\ref{eq:isothermal}) teaches us a few lessons about the degeneracies inherent in transmission spectra:
\begin{itemize}

\item The transit radius may be rewritten as $R = C + H \ln{\kappa}$, where $R_0$ and $P_0$ are absorbed into the constant $C$.  This implies that there is a degeneracy between $R_0$ and $P_0$.  This degeneracy was noticed numerically by \cite{bs12}.  In practice, $R_0$ may be matched to the measured white-light radius, but the value of $P_0$ is unknown and cannot be extracted from the data.

\item If $R_0$ is matched to the white-light radius, it means that it does not possess an exact value, but rather a range of values that is associated with the measured uncertainties in the white-light radius.  However, since the functional relationship between $R_0$ and $P_0$ is unknown, it is not apparent how one should adjust the value of $P_0$ as $R_0$ is varied.

\item The opacity may generally be written as $\kappa = \chi \kappa_0$, where $\chi$ is the mass mixing ratio\footnote{Relative abundance by mass.} (and not the volume mixing ratio\footnote{Relative abundance by number.}) and $\kappa_0$ is the opacity of a specific molecule (e.g., water).  However, $\chi$ can be absorbed into the constant $C$, which informs us that there is a degeneracy between $\chi$, $R_0$ and $P_0$.  In particular, it is $P_0\chi$, rather than $\chi$, that one really infers from a fit to data.

\item The pressure scale height controls the shape of the transmission spectrum, but not its normalization.  This implies that the temperature is degenerate with the normalization, which is controlled by $R_0$, $P_0$ and $\chi$.

\end{itemize}

If we specialize to WFC3 transmission spectra, then these degeneracies inform us that the water abundance\footnote{From this point on, we will use the terms ``abundance" and ``mixing ratio" interchangeably.  The latter may refer to either the volume or mass mixing ratio, depending upon context.} cannot be uniquely inferred from the data, and depend strongly on the assumed values of the reference transit radius and reference pressure.  Since the water mixing ratio appears in the logarithm and the reference transit radius does not, it follows that small variations in $R_0$ will lead to large variations in $\chi$ \citep{g14}.  We will demonstrate this point later during our analysis of WFC3 data of WASP-12b.

\subsection{Non-isothermal atmospheres}

\begin{figure}
\includegraphics[width=\columnwidth]{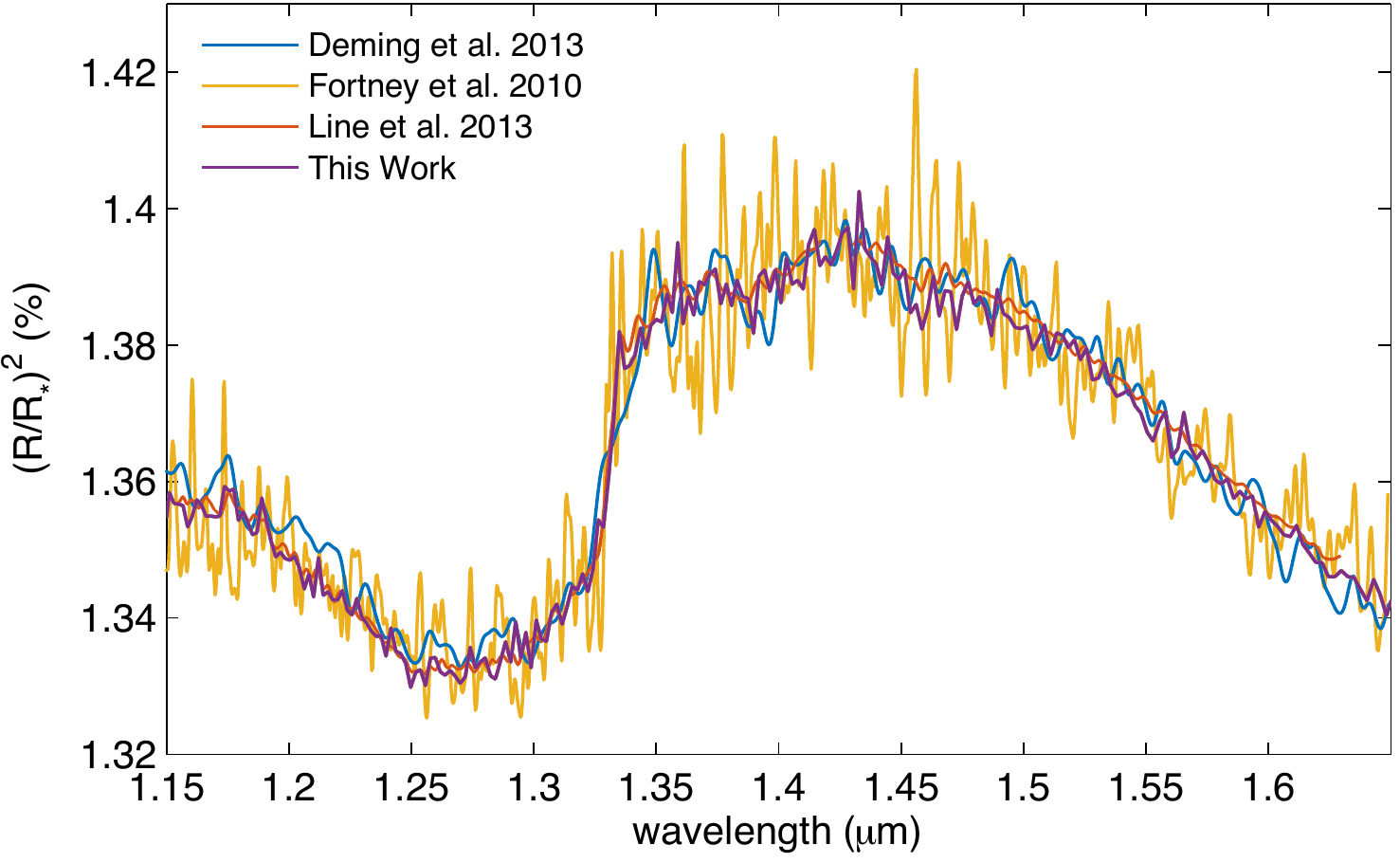}
\includegraphics[width=\columnwidth]{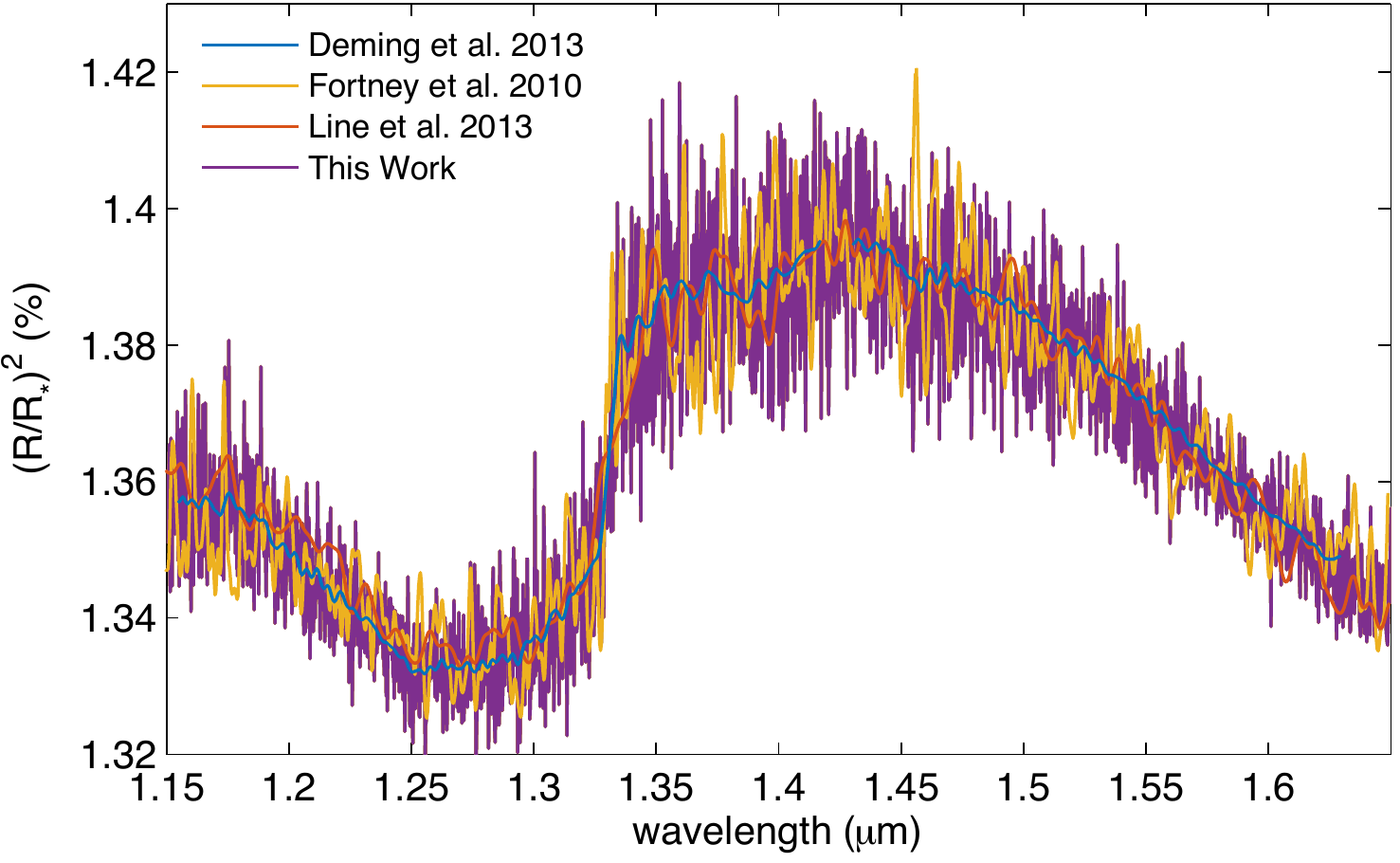}
\vspace{-0.1in}
\caption{Benchmarking our full numerical calculation of the transmission spectrum of HD 209458b (see text for details of input parameters) to those of Fortney et al., Deming et al., and Line et al.  We used a spectral resolution of $\sim 0.1$ cm$^{-1}$ for the opacity function to compute the model transmission spectrum.  In the top and bottom panels, we then binned down the model to resolutions of $\sim 10$ cm$^{-1}$ and $\sim 1$ cm$^{-1}$, respectively, to illustrate that minor discrepancies result from adopting different spectral resolutions for the water opacity.  (Discrepancies may also arise from the use of different spectroscopic line lists to construct the opacity function.)}
\label{fig:benchmark}
\end{figure}

\begin{figure*}
\includegraphics[width=\columnwidth]{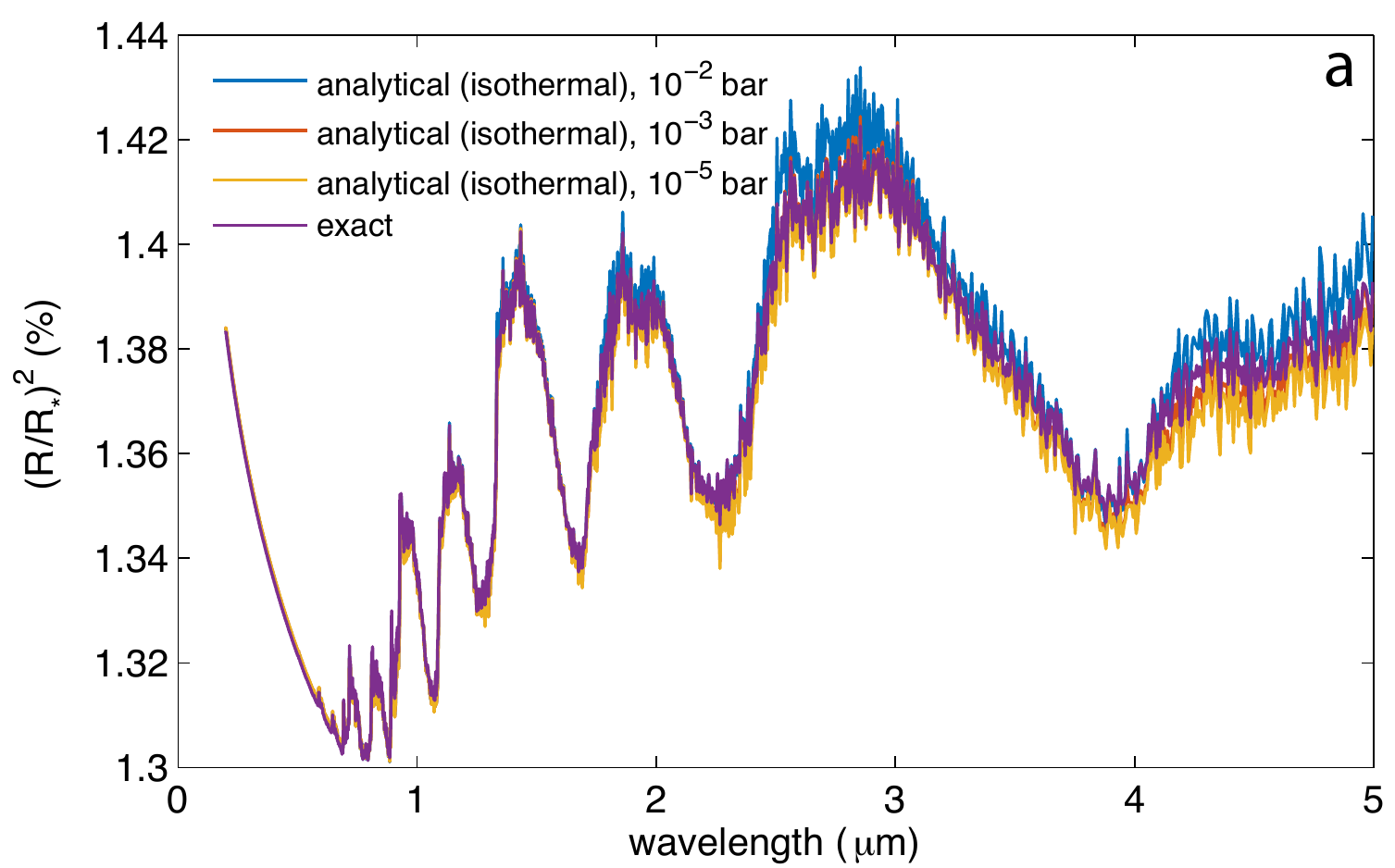}
\includegraphics[width=\columnwidth]{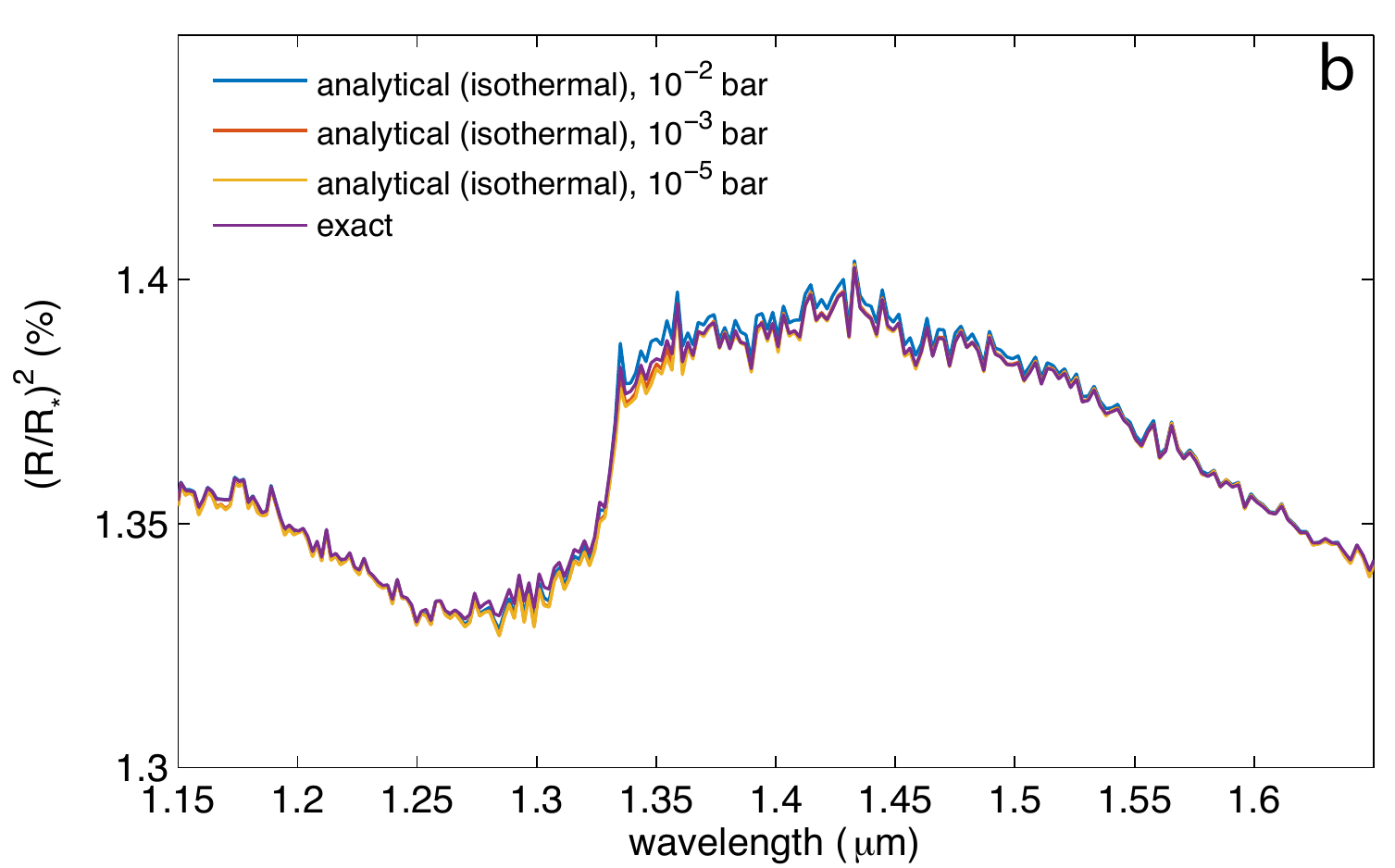}
\includegraphics[width=\columnwidth]{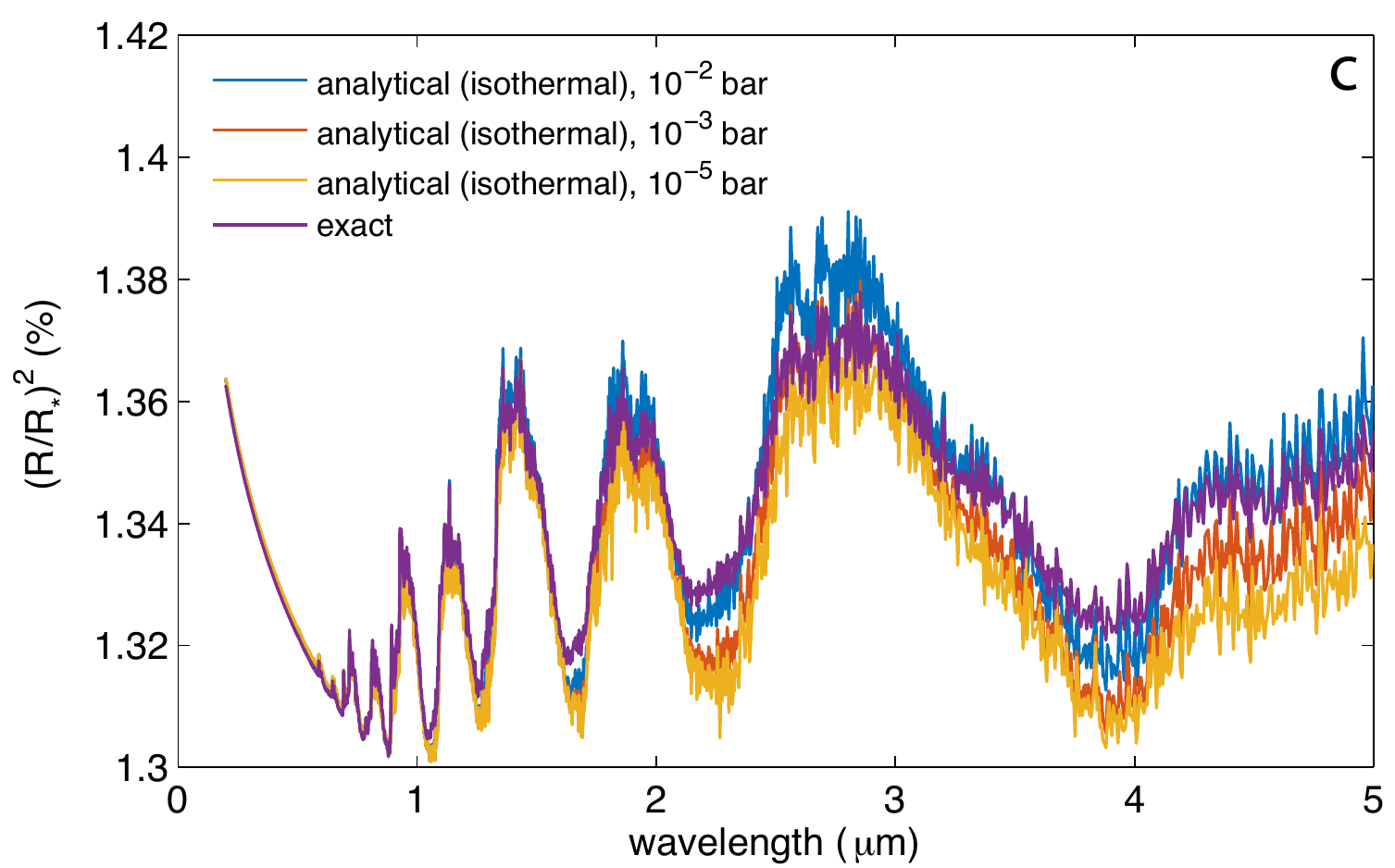}
\includegraphics[width=\columnwidth]{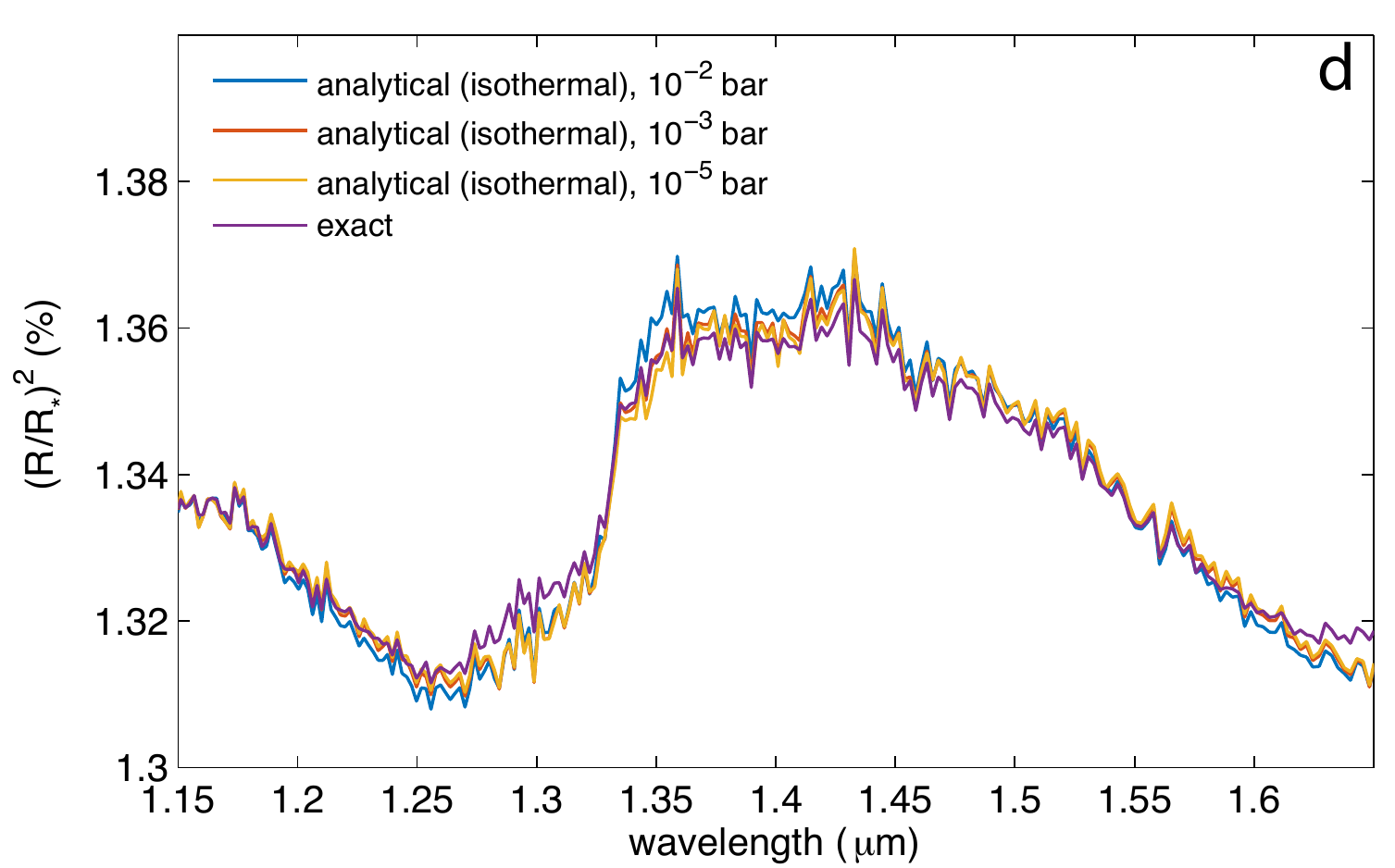}
\includegraphics[width=\columnwidth]{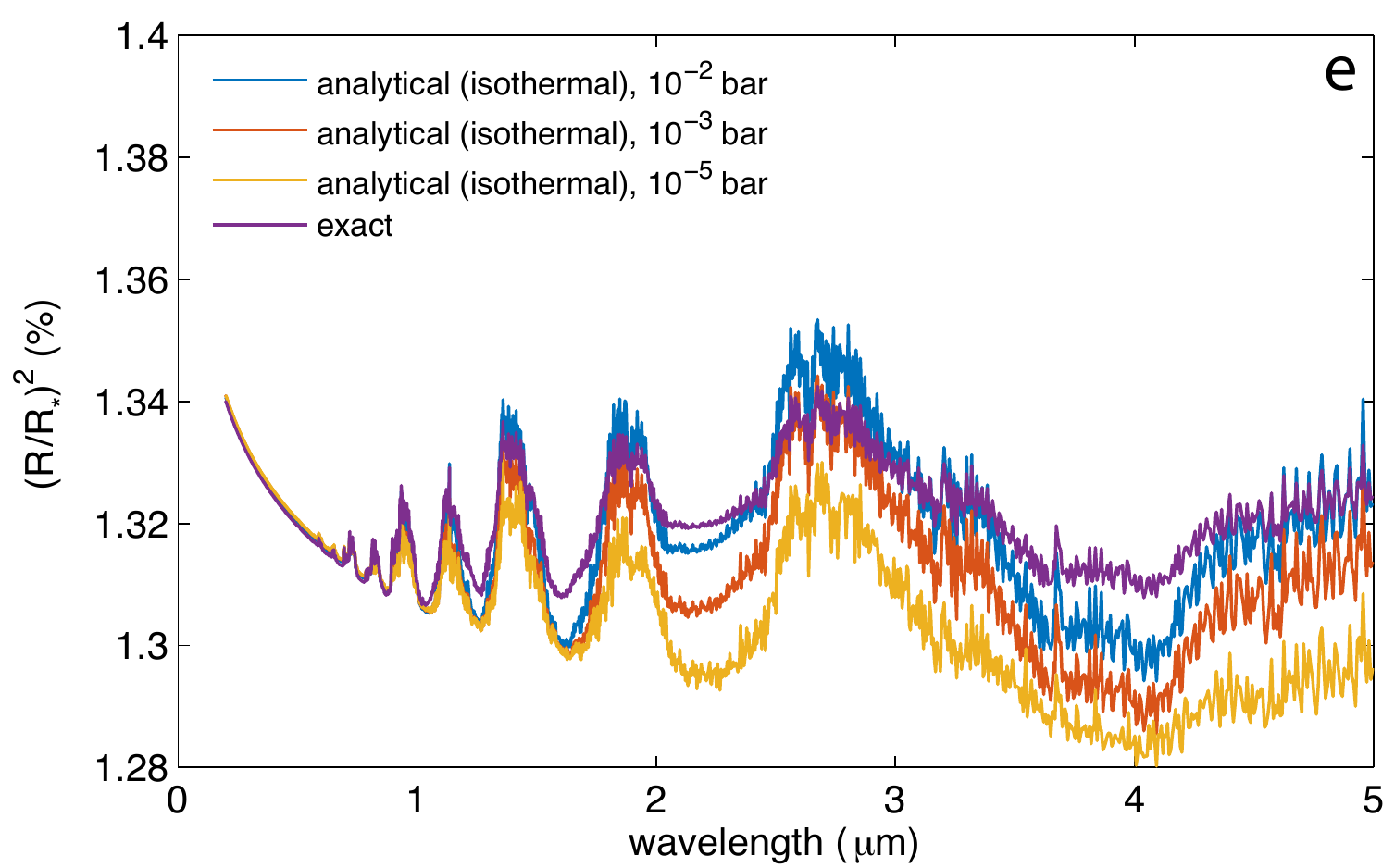}
\includegraphics[width=\columnwidth]{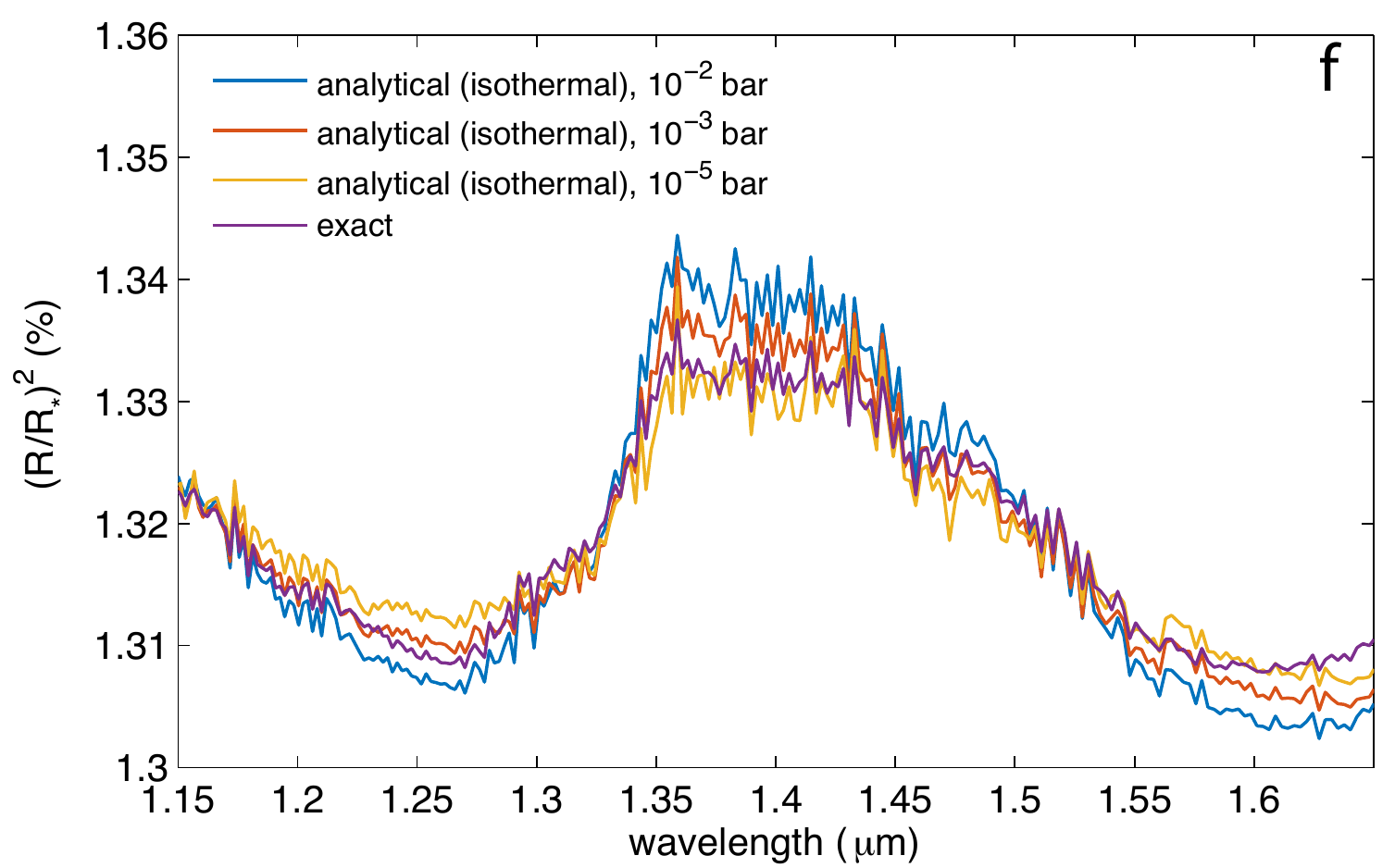}
\caption{Comparing our full numerical calculations of the transmission spectrum of an isothermal atmosphere to those computed using our isothermal, isobaric analytical formula.  The first, second and third rows are for $T=1500$ K, $T=1000$ K and $T=500$ K, respectively.  The left and right columns are for JWST-NIRSpec and HST-WFC3 wavelength coverage, respectively.  We used a spectral resolution $\sim 0.1$ cm$^{-1}$ for the opacity function and then binned the model down to a resolution $\sim 10$ cm$^{-1}$.  These calculations demonstrate that WFC3 transmission spectra may be approximated as being isobaric (constant pressure) in some instances.}
\label{fig:isothermal}
\end{figure*}

\begin{figure}
\includegraphics[width=\columnwidth]{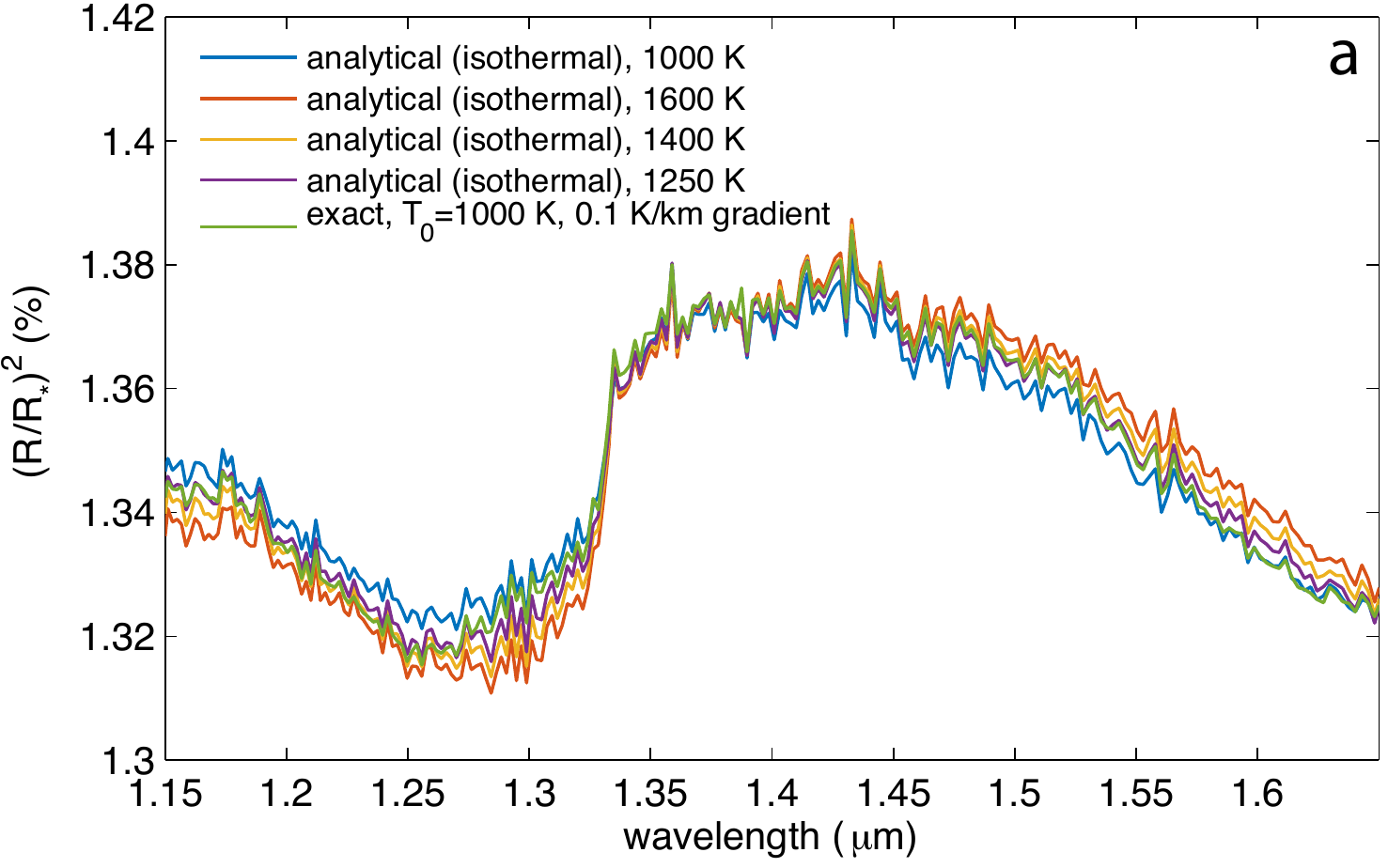}
\includegraphics[width=\columnwidth]{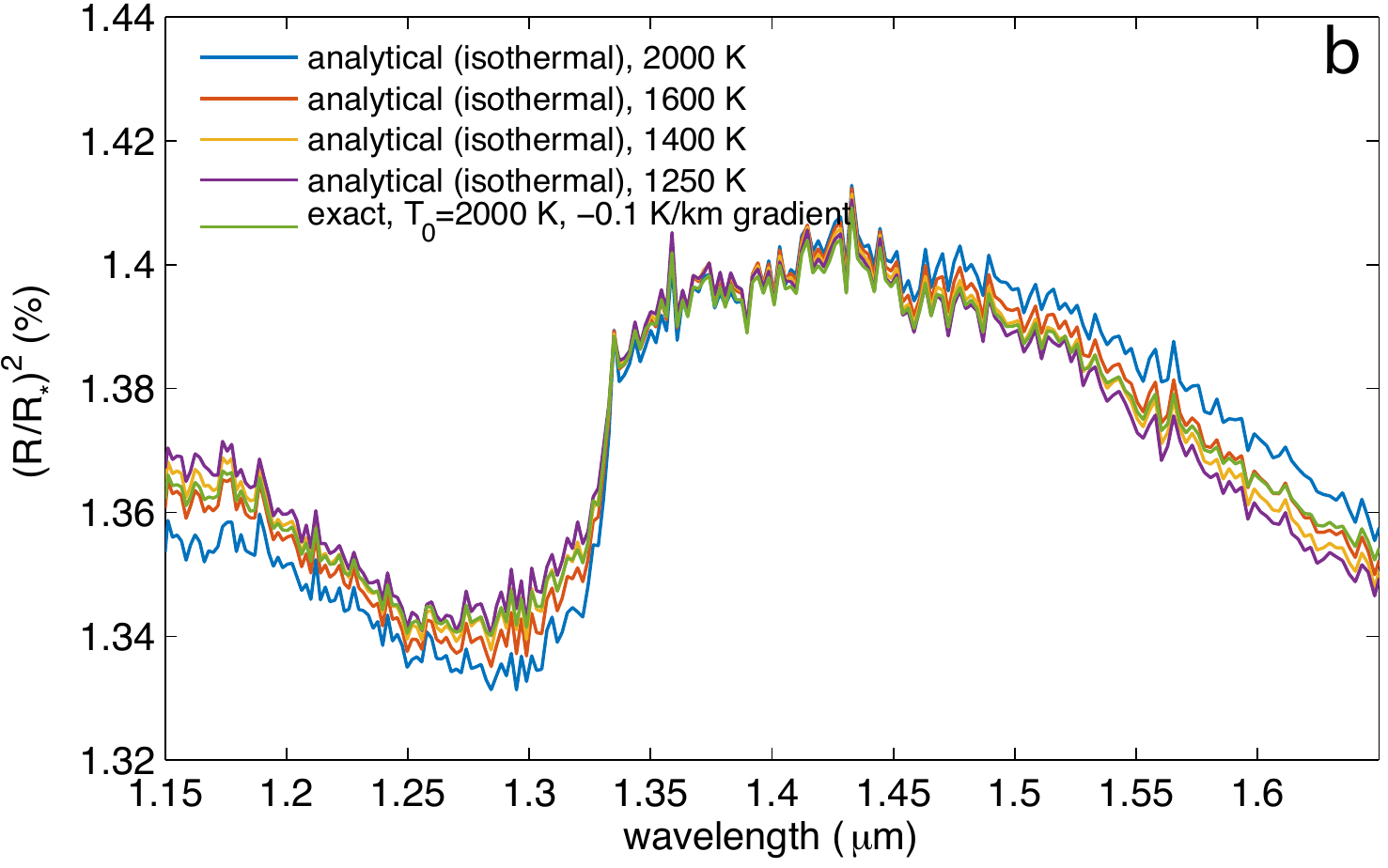}
\vspace{-0.1in}
\caption{Comparing full, non-isothermal numerical calculations to our isothermal, isobaric analytical formula.  The top and bottom panels show model atmospheres with positive and negative temperature gradients, respectively.  We used a spectral resolution $\sim 0.1$ cm$^{-1}$ for the opacity function and then binned the model down to a resolution $\sim 10$ cm$^{-1}$.  These calculations demonstrate that WFC3 transmission spectra may be accurately analyzed using an isothermal, isobaric model, which assumes a pressure of 1 mbar for the water opacities.  }
\vspace{-0.1in}
\label{fig:nonisothermal}
\end{figure}

Transmission spectra probe the atmosphere at high altitudes (low pressures), where the temperature-pressure profile is approximately linear with distance \citep{fortney08,fortney10}.  Furthermore, since transmission spectra are sensitive to only a limited range in pressure (Figure \ref{fig:schematic}), one may approximate the temperature-pressure profile as a constant with higher-order corrections.  For these reasons, we consider a non-isothermal model atmosphere with a temperature profile given by 
\begin{equation}
T = T_0 \pm T^\prime \left( r - R_0 \right).
\end{equation}
$T^\prime$ is the radial temperature gradient and is defined to be always a positive number.  Positive and negative signs correspond to temperature profiles with positive and negative gradients, respectively.  The quantity $T_0$ is the reference temperature corresponding to $R_0$ and $P_0$.  We consider this non-isothermal correction to the temperature to be small.  For example, we expect $T^\prime \sim 0.1$ K km$^{-1}$ \citep{huitson12,heng15}, $T_0 \sim 1000$ K and $H \sim 100$ km for hot Jupiters, which yields $T^\prime H / T_0 \sim 10^{-2}$.  We then take the simplification that the opacity or cross section may be evaluated at $T=T_0$.

There is a need to distinguish between the isothermal pressure scale height, $H=kT_0/mg$, and the non-isothermal pressure scale height,
\begin{equation}
H^\prime = \frac{T_0}{T^\prime}.
\end{equation}
The ratio of these two quantities is
\begin{equation}
b \equiv \frac{H^\prime}{H} = \frac{mg}{kT^\prime}.
\end{equation}
Here, $m$ is the mean molecular mass.  For hydrogen-dominated atmospheres with $T^\prime \sim 0.1$ K km$^{-1}$, we have $b \approx 30$.  For comparison, the Earth has $T^\prime \sim 1$ K km$^{-1}$.

By integrating the expression for hydrostatic balance, we obtain
\begin{equation}
n = n_0 \left( 1 \pm \frac{r-R_0}{H^\prime} \right)^{\mp b - 1},
\end{equation}
where the positive and negative signs associated with the $(r-R_0)/H^\prime$ term correspond to positive and negative temperature gradients, respectively.  The chord optical depth is
\begin{equation}
\tau = \zeta \pi n \sigma \sqrt{\frac{2 T_0 R_0}{T^\prime}}.
\end{equation}
The exact value of the dimensionless coefficient $\zeta$ depends on evaluating the integral \citep{heng15},
\begin{equation}
\int^{+\pi/2}_{-\pi/2} \left( \cos{x} \right)^N ~dx = \zeta \pi,
\end{equation}
where $N = 2b$ if the temperature profile has a positive gradient and $N=2b-1$ if it has a negative gradient.  In practice, the value of $\zeta$ ranges between 0.1 and 1 and only appears as $\ln{\zeta}$ in the expression for the transit radius, which suggests that it is sufficient to set it to $\zeta=0.5$.  To do better requires an iteration between the inferred value of $b$ and a recalculation of $\zeta$ until convergence attains.

Again, we need to evaluate the integral,
\begin{equation}
h = \frac{1}{R_0} \int^{+\infty}_{R_0} \left[ 1 - \exp{\left(-\tau\right)} \right] r ~dr.
\end{equation}
We find that an analytical solution obtains only if $r ~dr \propto d\tau/\tau$, which occurs when we assume $b \gg 1$.  This yields
\begin{equation}
r ~dr \approx -\frac{H^\prime R_0}{b \tau} \tau_0^{\pm 1/b} ~d\tau.
\end{equation}  
It follows that the transit radius for a non-isothermal atmosphere is
\begin{equation}
R = R_0 + H \tau_0^{\pm 1/b} \left( \gamma + \ln{\tau_0} + E_1 \right),
\label{eq:nonisothermal}
\end{equation}
where the reference optical depth is given by
\begin{equation}
\tau_0 = \frac{\zeta \pi P_0 \kappa}{g} \sqrt{\frac{2 R_0 b}{H}}.
\end{equation}
As before, the $E_1$ term may be dropped if $\tau_0 \gg 1$ is assumed.

By comparing equation (\ref{eq:nonisothermal}) to its isothermal counterpart in equation (\ref{eq:isothermal}), we see that the formulae for the transit radius are very similar in structure.  The reference transit radius and reference pressure are joined by $b$ in the logarithm.  The main effect of non-isothermal behavior is to introduce the $\tau_0^{\pm 1/b}$ correction factor next to the isothermal pressure scale height.  It accounts for an enhancement to the transit radius when the temperature gradient is positive, and a diminution when it is negative.  Since this factor is degenerate with $H$, it implies that the strength of spectral features in transmission spectra is controlled by a combination of temperature, surface gravity and mean molecular mass (which make up the isothermal pressure scale height), as well as a finite temperature gradient.  These quantities balance one another out and thus are degenerate with one another.  It suggests that the isothermal formula in equation (\ref{eq:isothermal}) should suffice for fitting data---a statement we will prove in the next section.

\section{Results}
\label{sect:results}

\subsection{Benchmarking and validation}

The results in this subsection justify the use of isothermal, isobaric model atmospheres, described by the simple analytical formula in equation (\ref{eq:isothermal}), to analyze WFC3 transmission spectra for hot exoplanetary atmospheres.

\subsubsection{Benchmarking to the isothermal, non-isobaric calculations of Deming, Fortney and Line}
\label{subsect:benchmark}

We first wish to demonstrate that we are able to perform full calculations of transmission spectra (with no approximations taken) correctly.  In their Figure 5, \cite{line13} previously published a transmission spectrum of HD 209458b with $R_0 = 1.25 ~R_{\rm J}$ (with $R_{\rm J}$ being the radius of Jupiter), $P_0 = 10$ bar, $R_\star = 1.148 ~R_\odot$ (with $R_\odot$ being the solar radius) and $g=10$ m s$^{-2}$.  The model atmosphere is assumed to be isothermal with $T=1500$ K and discretized into 90 layers between 0.1 nbar and 10 bar (equally spaced in the logarithm of pressure).  The volume mixing ratios of molecular hydrogen, helium and water are assumed to be 0.85, 0.15 and $4.5 \times 10^{-4}$, respectively, which translate into $m = 2.3 ~m_{\rm amu}$.  The quantity $m_{\rm amu} = 1.660539040 \times 10^{-24}$ g is the atomic mass unit.  In the same figure, \cite{line13} included the calculations of \cite{fortney10} and \cite{deming13}, and demonstrated that all three calculations match well.

In Figure \ref{fig:benchmark}, we repeat this calculation and compare it against the calculations of \cite{fortney10}, \cite{deming13} and \cite{line13}.  Generally, there is excellent agreement and the discrepancies are at the level of $\sim 0.1\%$ or less.  The discrepancies are due to the different spectral resolutions in the opacity function adopted by the different groups.  We conclude that our full numerical calculations of transmission spectra are accurate.

\subsubsection{Validating isothermal, isobaric formula against full numerical calculations (isothermal, non-isobaric)}

Now that we have validated our numerical approach, we compare calculations of the transmission spectrum of the same $T=1500$ K isothermal atmosphere to those computed using the isothermal formula in equation (\ref{eq:isothermal}).  There is an ambiguity in the isothermal formula in that it is not clear what to assume for the pressure $P$ in the opacity function $\kappa(\lambda,T,P)$.  Our intuition is that if the line peaks (rather than the line wings) are dominant as an opacity source, then the computed transmission spectrum should be insensitive to the value of $P$.  In other words, as long as pressure broadening is unimportant the value of $P$ should be irrelevant.  This insight should hold at high temperatures, but break down at low temperatures because of the increasing relative importance of the line wings as an opacity source due to pressure broadening.

In Figure \ref{fig:isothermal}, we validate this hypothesis by demonstrating that the computed transmission spectra for $T=1500$ K are almost identical for $P=0.01$ mbar versus $P=10$ mbar over the wavelength range probed by WFC3.  (For the rest of the study, we assume $P=1$ mbar.)  Even over the wavelength range probed by the \textit{Near Infrared Spectrograph} (NIRSpec) on the \textit{James Webb Space Telescope} (JWST), the agreement is rather remarkable.  Table \ref{tab:errors} lists the minimum, maximum and mean errors associated with comparing our analytical formula to the full numerical calculations.  Generally, the error worsens as the temperature decreases and the wavelength range increases, as seen for the 1000 K and 500 K case studies in Figure \ref{fig:isothermal}.  It is worth emphasizing that these discrepancies are \textit{not} due to the isothermal assumption (as the numerical model itself is isothermal).  Rather, the influence of the line wings, which depends on pressure, becomes stronger as the temperature becomes lower.  For example, we see in Figure \ref{fig:isothermal} that the use of the analytical formula becomes sensitive to the choice of pressure value for the 500 K case study.

\subsubsection{Validating isothermal, isobaric formula against full numerical calculations (non-isothermal, non-isobaric)}

As a final validation step, we wish to demonstrate that our isothermal, isobaric analytical formula does a decent job of matching full, non-isothermal, non-isobaric numerical calculations.  When using the analytical formula, we fix the pressure associated with the water opacities at 1 mbar.  We assume a typical temperature gradient of $T^\prime = 0.1$ K km$^{-1}$, which at the order-of-magnitude level is the value inferred from measurements of HD 189733b \citep{huitson12,heng15,wy15}.  We consider two representative cases: $T_0=1000$ K with a positive temperature gradient and $T_0=2000$ K with a negative temperature gradient.  In Figure \ref{fig:nonisothermal}, we compare the pair of non-isothermal calculations against a set of isothermal calculations done using the analytical formula.  We see that the isothermal models with temperatures of 1250 K and 1600 K provide satisfactory matches to the numerical calculations for the positive- and negative-gradient case studies, respectively.  If a higher level of accuracy is desired, then one may use the non-isothermal formula.

\subsection{Analyzing the WFC3 transmission spectrum of WASP-12b}

\begin{table*}
\centering
\caption{Errors associated with comparing full numerical calculations to analytical formula in equation (\ref{eq:isothermal})}
\label{tab:errors}
\begin{tabular}{lccccccc}
\hline
Figure & $T$ probed & minimum & minimum & maximum & maximum & mean & mean \\
\hline 
 & & relative (\%) & absolute (ppm) & relative (\%) & absolute (ppm) & relative (\%) & absolute (ppm) \\
\hline
\hline
 3a & 1500 K & 1.53$\times 10^{-4}$ &  1.89$\times 10^{-2}$ & 0.61 & 72.98 &  0.11 & 13.19 \\
  3b & 1500 K & 4.75$\times 10^{-5}$ &  5.69$\times 10^{-3}$ & 0.56 & 67.72 &  0.11 & 13.85\\
  3c & 1000 K & 4.20$\times 10^{-4}$ &  5.25$\times 10^{-2}$ & 1.38 & 167.00 &  0.29 & 35.34\\
  3d & 1000 K & 1.31$\times 10^{-4}$ &  1.58$\times 10^{-2}$ & 1.10 & 132.91 &  0.31 & 37.96\\
  3e & 500 K & 2.96$\times 10^{-4}$ &  3.61$\times 10^{-2}$ & 1.79 & 213.67 &  0.41 & 49.09\\
  3f & 500 K & 3.51$\times 10^{-5}$ &  4.24$\times 10^{-3}$ & 1.24 & 148.14 &  0.47 & 56.53\\
  4a & 1250 K & 1.84$\times 10^{-4}$ &  2.19$\times 10^{-2}$ & 0.83 & 100.80 &  0.23 & 28.02\\
  4b & 1600 K & 7.09$\times 10^{-5}$ &  8.79$\times 10^{-3}$ & 0.88 & 114.40 &  0.18 & 22.80\\
\hline
\end{tabular}\\
Note: when using the analytical formula, a pressure of 1 mbar was assumed for the water opacity.
\end{table*}

\subsubsection{Comparison to specific models to study trends (no data fitting performed)}

Now that we have demonstrated the accuracy of the isothermal, isobaric analytical formula for computing WFC3 transmission spectra, we return to the models presented in Figure \ref{fig:wasp12b}.  \cite{k15} have previously measured and interpreted the WFC3 transmission spectrum of the hot Jupiter WASP-12b.  They assumed the stellar radius to be $R_\star = 1.57 ~R_\odot$.  The reference transit radius was fixed at $R_0 = 1.79 ~R_{\rm J}$, while the reference pressure was assumed to be $P_0=10$ bar.  As a proof of concept,  we adopt the \cite{k15} values for $R_\star$ and $P_0$, but explore four models where we varied the value of $R_0$.  The values of $R_0$ adopted are listed in Table \ref{tab:wasp12b}.  Note that these values of $R_0$ are well within the uncertainties associated with the white-light radius of $1.79 \pm 0.09 ~R_{\rm J}$ measured by \cite{hebb09}.  \cite{k15} do not report the value of the surface gravity used, but we follow \cite{hebb09} and use $\log{g} = 2.99$ (cgs units) or $g = 977$ cm s$^{-2}$.

To compute model transmission spectra, we use the isothermal formula in equation (\ref{eq:isothermal}) and also the non-isothermal formula in equation (\ref{eq:nonisothermal}).  We add a constant opacity associated with clouds or aerosols to the total opacity, which is described by
\begin{equation}
\kappa = \frac{m_{\rm H_2O}}{m} X_{\rm H_2O} \kappa_{\rm H_2O} + \kappa_{\rm cloud},
\end{equation}
where $m_{\rm H_2O} = 18 ~m_{\rm amu}$ is the mass of the water molecule, $X_{\rm H_2O}=10^{-3}$ is the volume mixing ratio of water and $\kappa_{\rm H_2O}$ is the water opacity.  In other words, $X_{\rm H_2O} m_{\rm H_2O} / m$ is the mass mixing ratio.  We set $m=2.4 ~m_{\rm amu}$.  The values of $\kappa_{\rm cloud}$ and $T^\prime$ adopted are listed in Table \ref{tab:wasp12b}.  Our constant cloud opacity assumes that the (spherical) cloud particles have circumferences that are larger than the longest wavelength probed, as described by Mie theory (e.g., \citealt{p10}).  Generally, higher values of the cloud opacity are needed as the temperature increases (which increases the water opacity).

\begin{figure}
\includegraphics[width=\columnwidth]{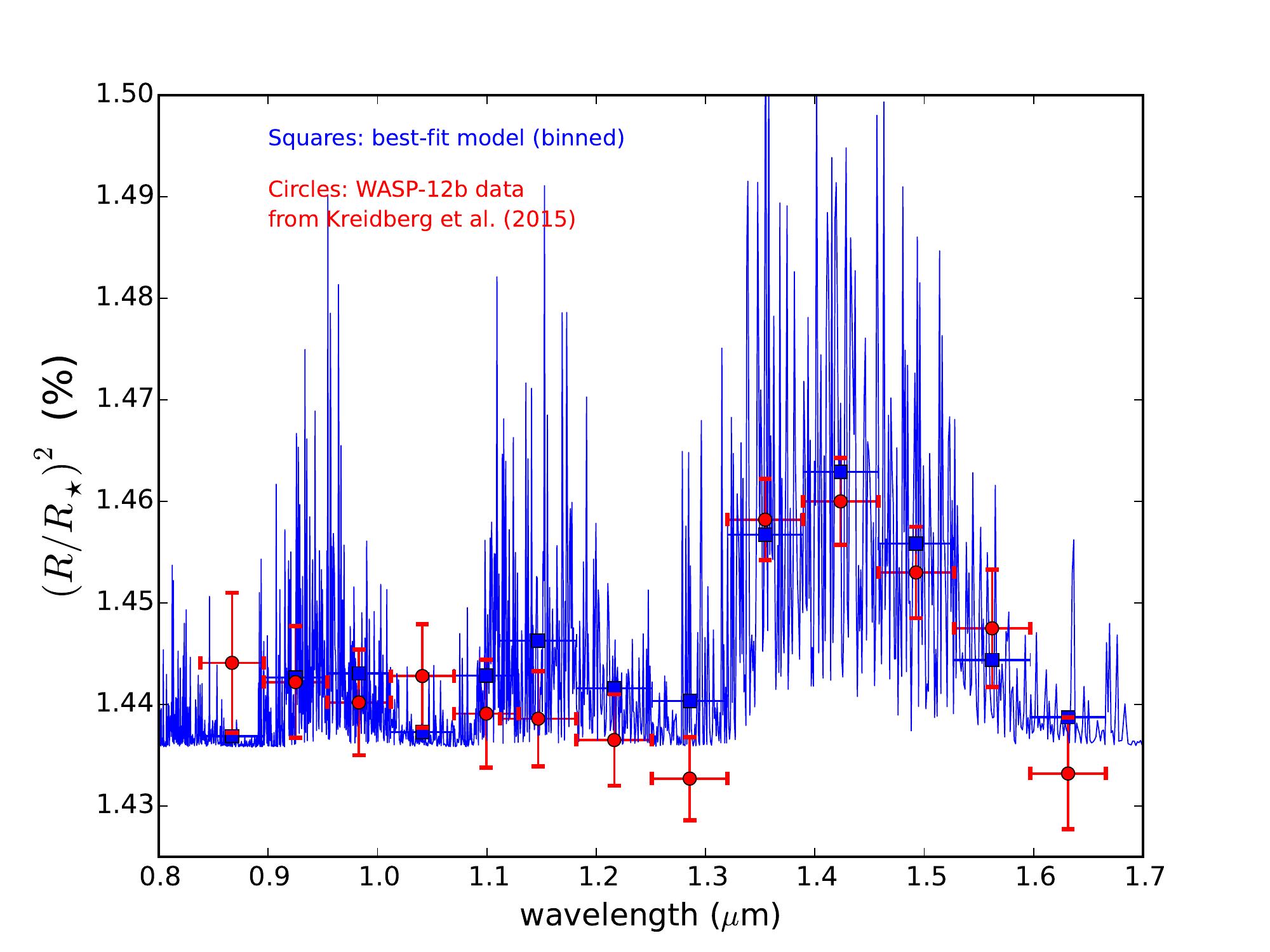}
\vspace{-0.1in}
\caption{Best-fit model using our analytical isothermal, isobaric formula to the measured WFC3 transmission spectrum of WASP-12b from Kreidberg et al. (2015).}
\vspace{-0.1in}
\label{fig:fit}
\end{figure}

\begin{figure}
\includegraphics[width=\columnwidth]{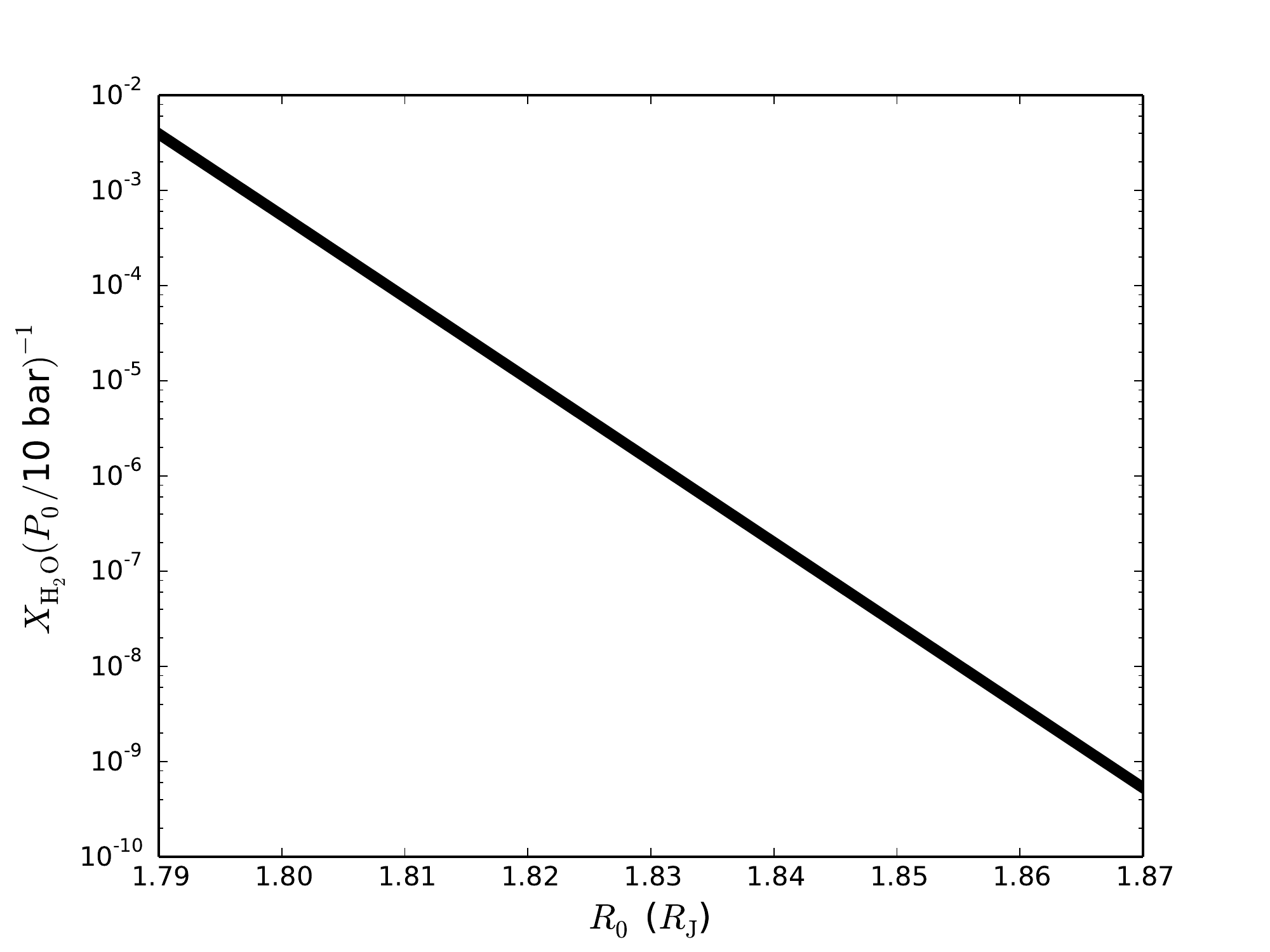}
\vspace{-0.1in}
\caption{Inferred water abundances, obtained from best fits to the measured WFC3 transmission spectrum of WASP-12b from Kreidberg et al. (2015), versus the assumed value of the reference transit radius.}
\vspace{-0.1in}
\label{fig:fit2}
\end{figure}

As already discussed in the beginning of the study, Figure \ref{fig:wasp12b} shows us that the temperature, temperature gradient and degree of cloudiness are not easily teased apart when interpreting the data.  This proof-of-concept comparison with the \cite{k15} data informs us that cloudfree models are ruled out, because it is difficult to match the relatively flat continuum blueward of the 1.4-$\mu$m water feature with cloudfree models, but it is harder to pin down the values of temperature and cloud opacity, and also if a temperature gradient is present.  Higher temperatures may be compensated by lower values of $R_0$.  The values of $R_0$ and $\kappa_{\rm cloud}$ may in turn be adjusted to compensate for each other.  The most discerning data points appear to be at 1.3 and 1.6 $\mu$m, just blueward and redward of the water absorption feature, respectively.  These sensitivity of these data points to temperature and cloudiness allow us to perform a fit to the data, provided we fix the values of $R_0$ and $P_0$.

\subsubsection{Fits to data: the $X_{\rm H_2O}$-$R_0$-$P_0$ degeneracy}

Figure \ref{fig:fit} shows our formal, three-parameter fit to the WFC transmission spectrum of WASP-12b measured by \cite{k15}, using the isothermal, isobaric formula in equation (\ref{eq:isothermal}).  The details of this procedure are described in the Appendix.  We first wish to demonstrate that our simpler approach yields the same answer as the full retrieval method of \cite{k15}.  By setting $R_0 = 1.79 ~R_{\rm J}$ and $P_0=10$ bar, we obtain $T=1020$ K and $X_{\rm H_2O} = 3.9 \times 10^{-3}$, which is consistent with the range of values reported in Table 4 of \cite{k15}.  The cloud opacity is $\kappa_{\rm cloud} = 3.5 \times 10^{-3}$ cm$^2$ g$^{-1}$.  \cite{k15} mention the use of a planetary radius scaling factor ``to account for uncertainty in the pressure level in the atmosphere at a given radius".  Their Figure 11 shows that this scale factor is $0.99^{+0.01}_{-0.02}$, essentially very close to unity.  Recall that variations in the reference pressure appear as $H \ln{P_0}$ in the formula for the transit radius, which implies that they are muted by the logarithm even if $P_0$ is varied by several orders of magnitude.  By contrast, variations in $R_0$ affect the transit radius linearly and produce large changes in the water mixing ratio.  It is unclear if this scaling factor accounts for the uncertainties in the white-light radius of $1.79 \pm 0.09 ~R_{\rm J}$ measured by \cite{hebb09}.

Next, we set $R_0 = 1.85 ~R_{\rm J}$, which is a 3.4\% increase from its previous value.  Remarkably, such a small increase in the reference transit radius leads to more than 5 orders of magnitude of decrease in the water mixing ratio obtained from the fit: $X_{\rm H_2O} = 2.8 \times 10^{-8}$.  The cloud opacity becomes $\kappa_{\rm cloud} = 2.5 \times 10^{-8}$ cm$^2$ g$^{-1}$.  Essentially, cloudiness and water abundance may be cancelled out by a larger normalization.  Furthermore, it is really $P_0 X_{\rm H_2O}$ that is the fitting parameter and not $X_{\rm H_2O}$ alone.  Any linear change in the reference pressure is compensated by the same factor in the water mixing ratio.  In Figure \ref{fig:fit2}, we show the values of $X_{\rm H_2O} (P_0/\mbox{10 bar})^{-1}$ obtained from fitting to the data as a function of the value of $R_0$ assumed.

A more detailed exploration of these degeneracies and the posterior distributions of parameters is beyond the scope of the present study and deferred to future work.

\subsection{An unresolved challenge: how do we relate $R_0$ and $P_0$ when analyzing data?}

An outstanding issue with setting $R_0$ to be the white-light radius is that it is inconsistent with the assumption of $\tau_0 \gg 1$.  At all wavelengths, a transmission spectrum is probing transit chords with optical depths $\sim 1$.  A white-light transit radius would correspond to $\tau_0 \sim 1$, not $\tau_0 \gg 1$.  This inconsistency may be alleviated by restoring the $E_1=E_1(\tau_0)$ term in equation (\ref{eq:isothermal}), yielding
\begin{equation}
R = R_0 + H \left[ \gamma + E_1 + \ln{\left( \frac{P_0 \kappa}{g} \sqrt{\frac{2 \pi R_0}{H}} \right)} \right].
\end{equation}
The preceding expression is now valid for all values of $\tau_0$, but the issue remains that the functional form of $R_0(P_0)$ is unknown.

Phenomenologically, one may fix $P_0$ and fit for the value of $R_0$.  The fit to the data will produce a solution for $R_0$, but it is unclear if this value of the reference transit radius corresponds correctly to the chosen value of $P_0$.  To give a concrete example, we return to the case of WASP-12b.  We use the white-light radius to set $R_0 = 1.79 \pm 0.09 ~R_{\rm J}$ \citep{hebb09}.  But what is the value of $P_0$?  The measured transmission spectrum has nothing to say on this issue, as one can only fit for the value of $P_0 X_{\rm H_2O}$ and not $X_{\rm H_2O}$ alone.

\textit{Ideally, we would like to fix $R_0$ to the white-light radius and use theory to inform us what $P_0$ is.}  For isothermal, isobaric atmospheres, we have \citep{heng16}
\begin{equation}
P_0 \sim 0.56 ~\frac{g}{\bar{\kappa}} \sqrt{\frac{H}{2 \pi R_0}},
\end{equation}
where we interpret $\bar{\kappa}$ as the geometric mean opacity within the white-light bandpass.  The preceding expression is correct only at the order-of-magnitude level, because of  the use of the geometric mean opacity.  But it would break the degeneracy associated with the reference transit radius and reference pressure.  Having determined the values of both $R_0$ and $P_0$, we may then perform a fit to the data to infer the water abundance.

Using $g = 977$ cm s$^{-2}$, $T = 1020$ K, $m=2.4 ~m_{\rm amu}$ and $R_0 = 1.79 ~R_{\rm J}$, we obtain $H = 362$ km and $P_0 \sim 1 \mbox{ mbar} ~(\bar{\kappa}/0.01 \mbox{ cm}^2 \mbox{ g}^{-1})^{-1}$.

\section{Discussion}
\label{sect:discussion}

\subsection{Implications and possible solutions}

There are several major implications of our findings.
\begin{itemize}

\item Beyond identifying the presence of water in the WFC3 data, our ability to accurately infer the water abundance is limited, due to transmission spectra lacking an absolute normalization.  Our ignorance of this absolute normalization is due to the unknown relationship between the reference transit radius and the reference pressure.

\item It has previously been proposed that the J band (1.22--1.30 $\mu$m) and water spectral feature (1.36--1.44 $\mu$m), located within WFC transmission spectra, serve as diagnostics for the degree of cloudiness in an exoplanetary atmosphere \citep{stevenson16}.  The lack of an absolute normalization and the degeneracies associated with temperature, a finite temperature gradient and the degree of cloudiness suggest that the true picture is more complicated and the interpretation is multi-dimensional.

\item We suggest an approximate relationship between the reference transit radius and reference pressure.  However, since it is correct only at the order-of-magnitude level, it implies that the inferred abundance of water is only accurate at the order-of-magnitude level.

\end{itemize}

Other solutions to break the degeneracy have been suggested.  If pressure broadening of the line wings is a major source of opacity, then it will break the insensitivity of the transmission spectrum to the pressure being sensed, as previously suggested by \cite{g14}, although this is only useful at low temperatures.  This behaviour is clearly seen for the 500 K case study in Figure \ref{fig:isothermal}.  The caveat is that the theory of pressure broadening is incomplete, and the choice of line-wing cutoff\footnote{In the current study, we adopt a cutoff of 25 cm$^{-1}$ for the water lines.} may impact the inference made \citep{sb07,gh15}.  A cleaner idea is to use the measured transit radius associated with the optical/visible Rayleigh slope to pin down the pressure being sensed \citep{g14}.  This approach is unambiguous only if the atmosphere is cloud-free, such that Rayleigh scattering may be robustly associated with the bulk gas in the atmosphere \citep{heng16}.  In this case, the opacity due to Rayleigh scattering and the atmospheric pressure (dominated by the bulk gas) are straightforwardly related.

Yet another solution is to perform a joint analysis of emission and transmission spectra, as the former is not subjected to the degeneracies discussed \citep{g14}.  However, it remains unclear if the atmospheric conditions probed in emission and transmission may be described by a single, one-dimensional model, as hot exoplanets are expected to be three-dimensional entities \citep{burrows10,hmp11,hs15}.

Generally, breaking this degeneracy in the face of additional degeneracies introduced by the presence of clouds and hazes is an outstanding problem.  For all of these reasons, we consider this challenge to be unresolved.

\subsection{Comparison to Griffith (2014)}

Taken at face value, some of the conclusions made by \cite{g14} are qualitatively similar to those of the current study.  As such, we will now perform a detailed comparison to, and analysis of, the study of \cite{g14}.

No formal derivation of the transit radius was performed in \cite{g14}.  In fact, \cite{g14} does not show an analytical expression for $R$ at all, which is a key difference between that and the current study.  Rather, the study of \cite{g14} relies on analytical scalings, rather than analytical derivations---the latter refers to starting from first principles, whereas the former uses heuristic arguments.  One consequence of this approach is that the origin of the relationship between the reference radius ($R_0$) and reference pressure ($P_0$) was not elucidated.  \cite{g14} asserts that $R_0$ is the transit radius at a pressure where the atmosphere is opaque, but our derivation shows that this is not a necessary condition, and that $R_0(P_0)$ is simply a mathematical constant of integration.  By retaining the extra term involving the exponential integral, it is possible to associate $R_0$ with any optical depth.

\cite{g14} states that a degeneracy arises from ``uncertainties in the assignment of the pressure level associated with the retrieved radius at unity optical depth".  This is consistent with our finding that transmission spectra are insensitive to the pressure probed.  However, this is a separate and distinct degeneracy from the unknown relationship between $R_0$ and $P_0$, since the reference radius is generally not located at an optical depth of unity.  In fact, the lack of a first-principles analytical formula for $R$ is probably the reason why \cite{g14} did not discover the degeneracy between $R_0$ and $\ln{(P_0 X)}$, where $X$ is the volume mixing ratio.  \cite{g14} discusses two models of an isothermal atmosphere with slightly different values of the reference radius: $R_0$ and $R_0^\prime$.  Using our analytical formula, we realize that equation (3.6) of \cite{g14} is actually a simplified version of a more general, first-principles expression,
\begin{equation}
\frac{P_0 X}{P_0^\prime X^\prime} = \exp{\left(- \frac{\Delta R_0}{H} \right)} ~\sqrt{\frac{R_0^\prime}{R_0}},
\end{equation}
where $\Delta R_0 \equiv R_0 - R_0^\prime$.  While $\sqrt{R_0^\prime / R_0} \approx 1$ is probably a reasonable assumption (and it is made implicitly by \citealt{g14}), the real degeneracy is not just between the two mixing ratios ($X$ and $X^\prime$), but rather between $P_0 X$ and $P_0^\prime X^\prime$.

There is a discrepancy between our derivation that the variation in radial distance, across the transit chord, is $\pi H/4$ and the claim of \cite{g14} that it is about $4H$.  Both claims are made for isothermal atmospheres.  \cite{g14} arrived at this estimate by using an approximate scaling law for $\delta R$, bounded by a pair of transmission\footnote{Defined by \cite{g14} as being $\exp{(-\tau)}$, where $\tau$ is the optical depth.} values.  By choosing these transmission values to be 0.05 and 0.95 (corresponding to optical depths of about 3 and 0.05, respectively), \cite{g14} estimated that $\delta R \approx 4 H$.  These are rather ad hoc choices.  For example, choosing 0.01 and 0.99 instead yields $\delta R \approx 6 H$.  Figure 6 of \cite{g14} does not actually validate these choices, because it simply plots the value of the transmission as a function of depth/pressure.  In our approach, we have used the result that there is an effective transit chord with a chord length of $\sqrt{2 \pi H R}$ \citep{fortney05,lec08,ds13,heng15}, and used geometry to arrive at $\delta R = \pi H/4$ (Figure \ref{fig:schematic}).   

Pedantically (and for completeness), we note that \cite{g14} (re)discovered the $\sqrt{2 \pi H R}$ expression for the transit chord without crediting \cite{fortney05}.  Furthermore, there is no discussion or derivation of the result that the effective chord optical depth is 0.56.

\subsection{Why were these degeneracies not uncovered in retrieval studies?}

At first sight, these degeneracies do not seem to exist in the traditional forward problem (e.g., \citealt{h01,fortney08,fortney10}).  Given a set of assumptions (e.g., identities and abundances of molecules, chemistry, irradiation conditions), one may construct a model atmosphere and trace lines-of-sight (chords) through it to obtain the transmission spectrum.  For a fixed set of parameter values, one may uniquely calculate the relationship between each transit chord, the transit radius and pressure, and also infer the reference pressure (given a choice of the reference transit radius).

A retrieval modeler might plausibly object by stating that he or she has set up a grid in pressure or distance and explored a suite of model atmospheres with different chemistries, temperature-pressure profiles, etc.  For each model in the suite, the pressure corresponding to each transit radius, as well as the reference transit radius and reference pressure, may be straightforwardly calculated.  However, recall that because transmission spectra are insensitive to the pressure probed (unless pressure broadening comes into play) there exist multiple families of model atmospheres where the grid in pressure may be shifted up or down with little to no effect on one's ability to fit the observed spectrum.  For each choice of grid, there is a different set of $R_0(P_0)$ relationships.  Model suites with a fixed choice of pressure grid would not fully explore the degeneracies unearthed in the present study.

\vspace{0.3in}
\noindent
\textit{We acknowledge partial financial support from the Center for Space and Habitability (CSH), the PlanetS National Center of Competence in Research (NCCR), the Swiss National Science Foundation and the MERAC Foundation.  We thank Brice-Olivier Demory, Sara Seager, Laura Kreidberg, Michael Line and a level-headed referee for constructive conversations.  We are grateful to Mark Marley for insightful feedback on an earlier version of the manuscript.}

\appendix

\section{Details of Model Fitting}

We perform a fit of equation (\ref{eq:isothermal}) to the binned data points listed in Table 3 of \cite{k15} by implementing the \texttt{curve\_fit} routine in \texttt{Python}.  We take the following approximation when computing the model transmission spectrum: we compute the water opacities (with a pressure fixed at 1 mbar) in each of the wavebands listed in Table 3 of \cite{k15}.  Within each wandband, we compute the \textit{geometric mean} of the opacities.  We then use these mean opacities to compute the model transmission spectrum, which we fit to the binned data.  We check that our approach is sound by using the values of the fitting parameters to compute the transmission spectrum at full spectral resolution ($\sim 0.1$ cm$^{-1}$; shown as the curve in Figure \ref{fig:fit}), which we then bin down (shown as the squares in Figure \ref{fig:fit}) and compare to the data points (shown as the circles in Figure \ref{fig:fit}).  When this approach is repeated with the arithmetic mean of the opacities, it is apparent that the binned model does not match the data (not shown).  Physically, we expect the geometric mean to be a fair averaging of the opacities, since one is dealing with values that span many orders of magnitude.

\bsp	
\label{lastpage}
\end{document}